\documentclass[prl,twocolumn,superscriptaddress,nobibnotes]{revtex4-2}
\usepackage{graphicx}
\usepackage{amssymb}
\usepackage{amsmath}
\usepackage{bm}
\usepackage{color}
\usepackage{float}
\usepackage{setspace}
\usepackage{physics}
\usepackage{subfigure}
\usepackage{lineno}

\definecolor{orange}{RGB}{255,127,0}

\begin{document}
\title{Manipulating polariton condensates by Rashba-Dresselhaus \\ coupling at room temperature}

\author{Yao Li}
\affiliation{Tianjin Key Laboratory of Molecular Optoelectronic Science, Institute of Molecular Plus, Department of Physics, School of Science, Tianjin University, Tianjin 300072, China} 

\author{Xuekai Ma}
\affiliation{Department of Physics and Center for Optoelectronics and Photonics Paderborn (CeOPP), Universit\"{a}t Paderborn, Warburger Strasse 100, 33098 Paderborn, Germany}

\author{Xiaokun Zhai}
\affiliation{Tianjin Key Laboratory of Molecular Optoelectronic Science, Institute of Molecular Plus, Department of Physics, School of Science, Tianjin University, Tianjin 300072, China} 

\author{Meini Gao}
\affiliation{Tianjin Key Laboratory of Low Dimensional Materials Physics and Preparing Technology, School of Science, Tianjin University, Tianjin 300072, China}

\author{Haitao Dai}
\affiliation{Tianjin Key Laboratory of Low Dimensional Materials Physics and Preparing Technology, School of Science, Tianjin University, Tianjin 300072, China}

\author{Stefan Schumacher}
\affiliation{Department of Physics and Center for Optoelectronics and Photonics Paderborn (CeOPP), Universit\"{a}t Paderborn, Warburger Strasse 100, 33098 Paderborn, Germany}
\affiliation{Wyant College of Optical Sciences, University of Arizona, Tucson, AZ 85721, USA}

\author{Tingge Gao}
\affiliation{Tianjin Key Laboratory of Molecular Optoelectronic Science, Institute of Molecular Plus, Department of Physics, School of Science, Tianjin University, Tianjin 300072, China}

\begin{abstract}
\textbf{{Spin-orbit coupling plays an important role in the spin Hall effect, in topological insulators, and Bose-Einstein condensates with spin-orbit coupling show remarkable quantum phase transition. In this work we control an exciton polariton condensate -- a macroscopically coherent state of hybrid light and matter excitations -- by virtue of the Rashba-Dresselhaus (RD) spin-orbit coupling. This is achieved in a liquid-crystal filled microcavity where CsPbBr$_3$ perovskite microplates act as the gain material at room temperature. Specifically, we realize an artificial gauge field acting on the CsPbBr$_3$ exciton polariton condensate, splitting the condensate fractions with opposite spins in both momentum and real space. Besides the ground states, higher-order discrete polariton modes can also be split by the RD effect. Our work paves the way to manipulate  exciton polariton condensates with a synthetic gauge field based on the RD spin-orbit coupling at room temperature.}}
\end{abstract}

\maketitle

\section*{I\lowercase{ntroduction}}
The spin-orbit coupling describes the interaction between the spin and orbital degrees of freedom of a particle \cite{1-SOI review, 2-SOI book}. It leads to a plethora of physical phenomena such as the spin Hall effect \cite{1-SOI review,5-Kato Scienece,6-Spinhalleffect2} and topological insulators \cite{7-topological insulator1, 8-topological insulator2, 9-topological insulator3}. Dresselhaus \cite{3-Dresselhaus} and Rashba \cite{4-Rashba} proposed electronic spin-orbit coupling in noncentrosymmetric zinc-blende or wurtzite semiconductors. The engineering of the spin-orbit coupling in solids is usually limited by the growth or fabrication process of the materials such as high quality semiconductors. Thus an adjustable platform, where the spin-orbit coupling like the Rashba or Dresselhaus coefficient can be precisely manipulated, is highly sought after. Cold atom condensates offer the possibility to tune the spin-orbit coupling precisely using lasers to engineer the fine structure of the system states. For example, in the Bose-Einstein condensation (BEC) of cold atoms, the spin-orbit coupling provides the possibility to investigate quantum many-body phase transitions from a spatially spin-mixed state to a phase-separated state \cite{10-SOI atom BEC, 11-Liu X J SOI}. A stripe phase with supersolid properties was also demonstrated in a cold atom condensate \cite{Wolfgang Ketterle}. However, the ultra-low temperature needed to realize BECs of atoms limits the applicability and fundamental investigation of the spin-orbit coupling influencing bosonic condensates. 
Exciton polaritons, which form due to the strong coupling between excitons in quantum wells and cavity photons in a planar microcavity, can undergo a similar condensation process as the cold atoms, albeit in an inherently non-equilibrium state. Being a composite boson with small effective mass, an exciton polariton can condense at much higher temperature or even up to room temperature \cite{12-polariton BEC1, 13-polariton BEC2, 14-polariton BEC3}, so it can provide an alternative to the study of cold atom condensates with spin-orbit coupling \cite{polariton SOI} or synthetic gauge potentials \cite{non-Abelian gauge field} which can be electrically tuned \cite{ETH} without complicated laser cooling. Exciton polaritons can inherit the spin-orbit coupling of light \cite{15-SOI of light} from the cavity photon component that can be controlled by the detuning of the microcavity. The spin-orbit coupling of light results in the quantum spin Hall effect of light \cite{16-Bliokh Science}, spin Hall effect of light \cite{17-Geometrodynamics SOI}, and a non-cyclic optical geometric phase in a rolled-up microcavity \cite{18-SOI in asymmetric microcavity}. In a liquid crystal (LC) filled Fabry-Perot microcavity, a synthetic spin-orbit interaction similar to the spin-orbit interaction described by the original RD Hamiltonian has been realized at room temperature \cite{19-liquid crystal_science, 20}. One key point in that work is that a voltage can control the orientation of the molecular director of the LC molecules, which leads to an active modulation of the anisotropic refractive index of the cavity layer. 

Utilizing this concept in a system in which bosonic condensation with adjustable spin-orbit coupling can be investigated at room temperature requires an active material with robust excitonic excitations that strongly couple to the cavity mode with tunable RD coefficients. Also the exciton polariton condensation is needed which can be modulated by external means. Among the semiconductors that can sustain exciton polariton condensates at room temperature, the perovskite CsPbBr$_3$ has attracted considerable attention due to its supreme photonic properties. Inserted into a microcavity, CsPbBr$_3$ offers an excellent platform to investigate strong and novel light-matter interaction phenomena, such as exciton polariton condensation in a waveguide or a lattice \cite{20-xiong lattice, 21-xiong condensation} at room temperature. The anisotropy of the perovskite can be used to realize an effective Rashba Hamiltonian below condensation threshold \cite{22-Yang-Mills,28}, tune the Berry curvature \cite{tune berry curvature}, or observe a non-Hermitian topological invariant \cite{xiong non hermitian, topological phase}. However, the effects of RD spin-orbit coupling on exciton polariton condensates at room temperature remains unexplored. With the strong nonlinearity, RD spin-orbit coupled polariton condensates open the door to study intriguing spin dynamics, instability, vortices, many-body nonlinear physics \cite{Zhai hui}, and novel multistability phenomena with spin-orbit coupling.

In the present work we demonstrate the RD effect on the exciton polariton condensate formed in a planar cavity with CsPbBr$_3$ microplates as active material and filled with a LC layer. This LC microcavity \cite{19-liquid crystal_science, 23-liquid crystal Light} offers to tune the RD coefficients and realize a synthetic gauge field by applying a voltage across the microcavity. We insert CsPbBr$_3$ microplates into the cavity layer and demonstrate the exciton polariton condensation at room temperature. With the applied external voltage the polariton condensate can be actively manipulated. More importantly, the applied voltage across the microcavity tunes the exciton polariton condensate by the RD effect which acts as a synthetic gauge field and splits the exciton polaritons having opposite spins in real space and momentum space. Our work paves the way to explore spin orbit coupling induced quantum phase transitions based on exciton polariton condensates at room temperature.  

\begin{figure}[t]
	\centering
	\includegraphics[width=7.5 cm]{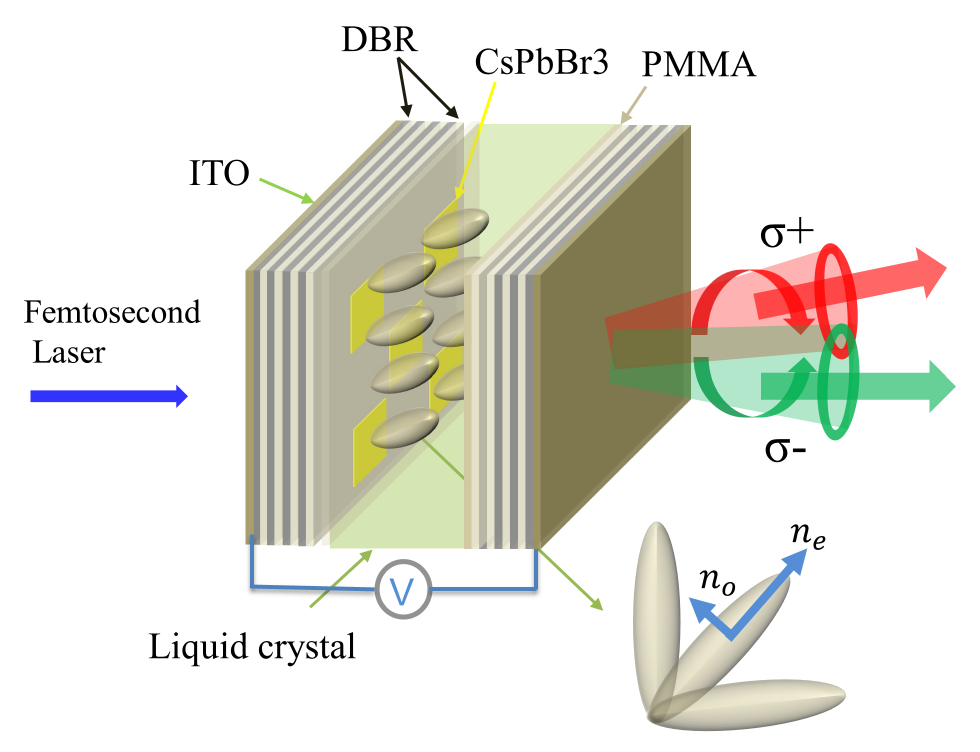} 
	\caption{\textbf{Schematics of the microcavity.} Schematic diagram of the liquid crystal microcavity structure with CsPbBr$_3$ inserted as the active layer. The anisotropy of the liquid crystal can induce spin orbit coupling of the polariton modes and split the polaritons with opposite spin, which is measured in the circularly polarized light emitted from the microcavity. } 
\end{figure}


\section*{R\lowercase{esults}}

In the LC microcavity where CsPbBr$_3$ microplates are inserted, on the one hand, cavity photons can strongly couple with excitons in the CsPbBr$_3$ layer. On the other hand, cavity photon modes can be tuned by electrically controlling the crystal molecules' orientation, giving rise to the RD effect. When the two effects occur simultaneously in this system, the system's Hamiltonian ($4\times4$) in the exciton-photon circularly polarized basis reads

\begin{equation}
H({\bf{k}})=
\begin{bmatrix}
E_\textup{x} & 0 & \Omega & 0 \\
0 & E_\textup{x} & 0 & \Omega \\
\Omega & 0 & E_\textup{p}+\dfrac{\hbar^2{\bf{k^2}}}{2m}+{\xi}k_{x} & \beta_{0}+\beta_{1}{\bf{k^2}}e^{2i\varphi} \\
0 & \Omega & \beta_{0}+\beta_{1}{\bf{k^2}}e^{-2i\varphi} & E_\textup{p}+\dfrac{\hbar^2{\bf{k^2}}}{2m} -{\xi}k_{x}\\
\end{bmatrix}.
\end{equation}

Here, $E_\textup{x}$ is the exciton energy which is the same for different spins. $E_\textup{p}$ is the ground state energy of cavity photon modes with effective mass $m$. The two cavity photon modes are perpendicularly linearly polarized and have different parity. The former leads to that they can strongly couple only to their corresponding excitons with the coupling strength $\Omega$ and the latter results in the RD effect with the spin-splitting strength $\xi$ along $k_x$ direction. $\beta_{0}$ denotes the XY splitting. $\beta_{1}$ represents the TE-TM splitting and $\varphi\in[0\ 2\pi]$ is the polar angle. The composite polariton particles inherit the RD interaction from the cavity photon component. The LC molecules can be rotated by the external voltage, which modulates the anisoptropic refractive index distribution within the microcavity. The changes of the refractive indices alter only one of the horizontally linearly polarized cavity photon modes, such that two lower-polariton (LP) branches can approach each other and consequently are shifted along $k_x$ direction, depending on the polarization, when they are brought into resonance.

\begin{figure}[t]
	\centering
	\includegraphics[width=8.5 cm]{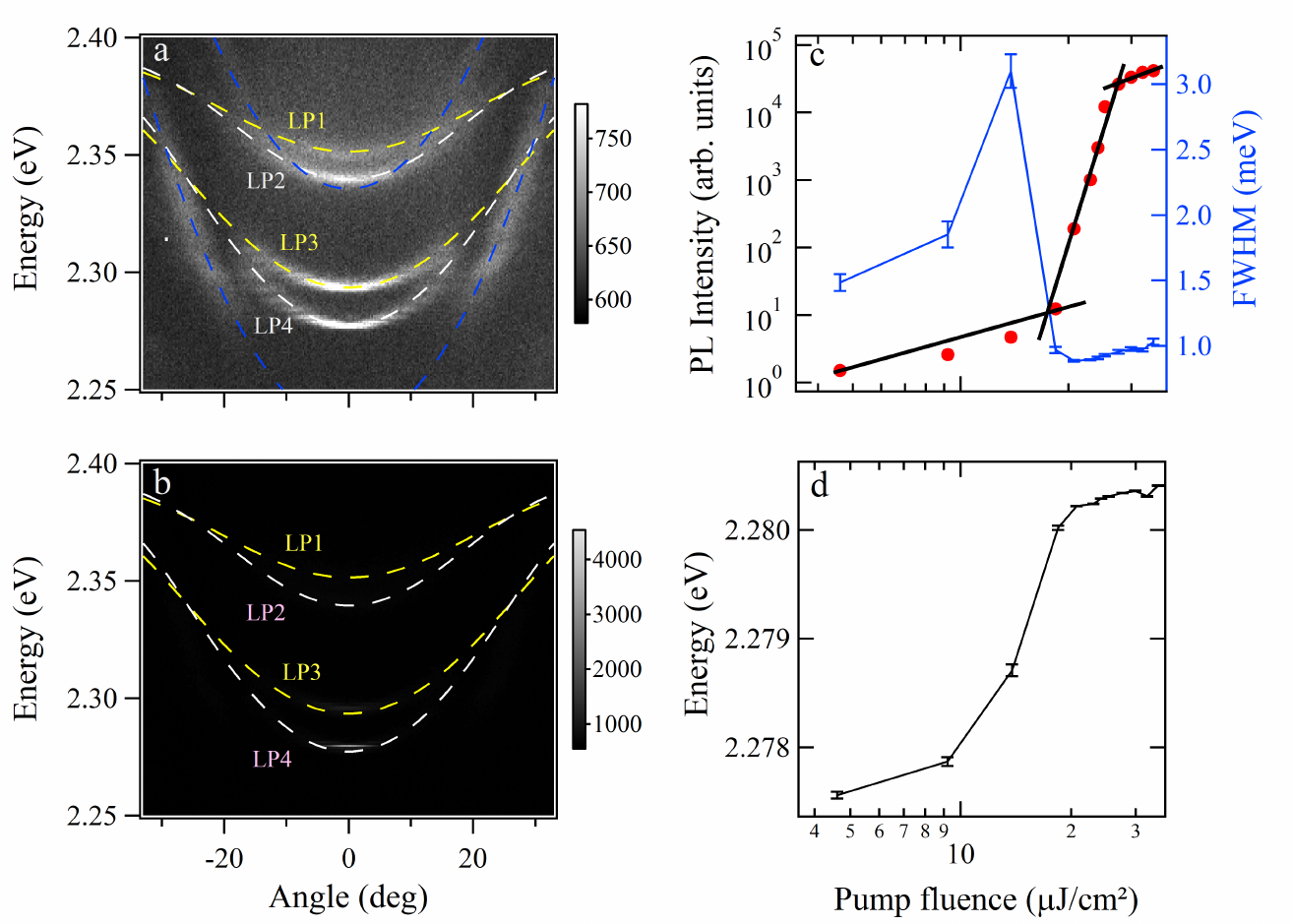} 
	\caption{\textbf{Polariton condensation in the LC microcavity.} (a) The photoluminescence dispersion of the microcavity below the threshold and (b) the condensate spectrum above the threshold for voltage of 2 V. (c) Integrated intensity and linewidth as  function of the pump fluence. (d) Polariton energy peak position as function of the pump fluence. \textcolor{blue}{The error bars in (c) are deviations of the estimated linewidth of polariton modes. And the error bars in (d) are deviations of the estimated energy of polariton modes.} The blue dashed lines in (a) are a guide for the eye to show the modes with large curvature. \textcolor{blue}{LP1-LP4 denote low branch polaritons 1-4.}}
	
\end{figure}

Firstly, we design the microcavity by calculating both the reflectivity and the electric field distribution of the structure using the transfer matrix method (the simulation results are shown in \textcolor{blue}{Supplementary Fig. 1}). The strong exciton-photon coupling can be realized thanks to the giant binding energy and oscillator strength of the excitons in the CsPbBr$_3$ microplate which is placed at the local maximum of the electric field inside the microcavity (c.f. \textcolor{blue}{Supplementary Fig. 1(b)}). In the experiments, the CsPbBr$_3$ microplate was grown by the CVD method (the details can be found in the Methods section), and then it was transferred onto the bottom DBR with the center wavelength of the stop band at 530 nm. The microcavity was fabricated by pasting the top DBR afterwards using adhesive. Finally, the LC is filled into the empty cavity layer to form the microcavity (the details of the microcavity can be found in the Methods section). The schematics of the whole structure of the sample is illustrated in Fig. 1.

We measure the dispersion of the microcavity for a voltage of 2 V by home-made momentum-space spectroscopy (see \textcolor{blue}{Supplementary Fig. 2 in the SI}). The excitation source is a femtosecond laser with the wavelength of around 400 nm and spot size of around 58 $\mu$m with Gaussian \textcolor{blue}{line profile}. The size of the CsPbBr$_3$ microplate we chose in the experiments is around 10 $\mu$m, which can be covered by the pump laser spot. The photoluminescence dispersion of the microcavity below the threshold is shown in Fig. 2(a), where several sets of modes are observed below the exciton resonance at 2.406 eV. We note that there exist some dispersions with much larger curvature, which come from the area outside the CsPbBr$_3$ microplate \cite{25-xiong steep dispersion} and do not affect the main results in the current experiments. The strong coupling is confirmed by the flatness of the bands at larger momenta (\textcolor{blue}{Supplementary Fig. 3(c)}). We note that the reflectivity of the polaritons is not visible due to high reflectivity of the DBRs at the stop band. Each set of the modes contains two lower branch polaritons (LP) due to the anisotropy of the microcavity induced by the LC cavity layer. Due to the finite size of the CsPbBr$_3$ microplate, the dispersions become discrete at larger momenta. The appearance of multiple sets of polaritons is due to the microcavity which sustains multiple cavity photon modes that can strongly couple with the near-resonant excitons. The curvature of the lower branch polariton modes is increased when the detuning between the cavity photon modes and the exciton resonance of the CsPbBr$_3$ microplate is smaller (more negative). We note that similar multiple polariton branches are also observed in a bare perovskite crystal \cite{24-sanvetto bare perovskite}. 

Within each set of modes in Fig. 2(a), the polarization of the two LP branches is perpendicular to each other: LP1 and LP3 are horizontally linearly polarized, while LP2 and LP4 are vertically linearly polarized. The parity of LP1 and LP2 are the same, and they are opposite to LP3 and LP4 which share the same parity too. Fig. 3(a-c) confirm the existence of the XY splitting due to the anisotropy of the microcavity. We note that the anisotropy of the CsPbBr$_3$ microplate itself can be neglected due to the very small thickness of around 200 nm. The polariton branches are fitted by coupled oscillator models and one can estimate the Rabi splitting of the polaritons within the microcavity to be around 90 meV for LP1-LP4. The exciton and cavity photon fractions of the multi-branch polaritons can be obtained by calculating the Hopfield coefficients (see \textcolor{blue}{Supplementary Fig. 4 in the SI}). It shows that the cavity photon components in the LP3 and LP4 are much higher than the LP1 and LP2 branches due to the more negative detuning, thus the dispersion curvature is larger as shown in Fig. 2(a).

Increasing the pump fluence to around 18.4 $\mu$J/cm$^2$ leads to the condensation of exciton polaritons, which can be seen in Fig. 2(b). The polaritons macroscopically occupy the ground state of the lower branch LP4 with a superlinear increase of the integrated intensity against the pumping flux, the reduction of the linewidth at the threshold, and the continuous blueshift of the energy peaks (Fig. 2(c) and Fig. 2(d)). To check whether the condensation is in the strong coupling regime, we \textcolor{blue}{replot} the data in Fig. 2(b) using a log scale (\textcolor{blue}{Supplementary Fig. 3(b) in the SI}). It shows that the dispersion at larger momenta (much weaker above the threshold) is the same as that below the threshold, evidencing the polariton condensation.  In real space the polariton condensate is localized near the center of the pump spot with the size of around 5 $\mu$m and the position of the localization depends on the inhomogeneity of the perovskites. To make sure whether macroscopic coherence develops or not above the threshold, we built a Michelson interferometer and interfered the real space images of the polariton condensate with itself rotated by 180 degrees.  Clear interference fringes are observed above threshold, whereas they are absent below the threshold (see \textcolor{blue}{Supplementary Fig. 5}). Further increasing the pumping fluence leads to polariton condensation at both LP3 and LP4 branches, as shown in \textcolor{blue}{Supplementary Fig. 6}. The linear polarization of the two polariton condensates with different energies are orthogonal to each other, which is the same as the result below the threshold in Fig. 2(a).

\begin{figure}
	\centering
	
	\includegraphics[width=\linewidth]{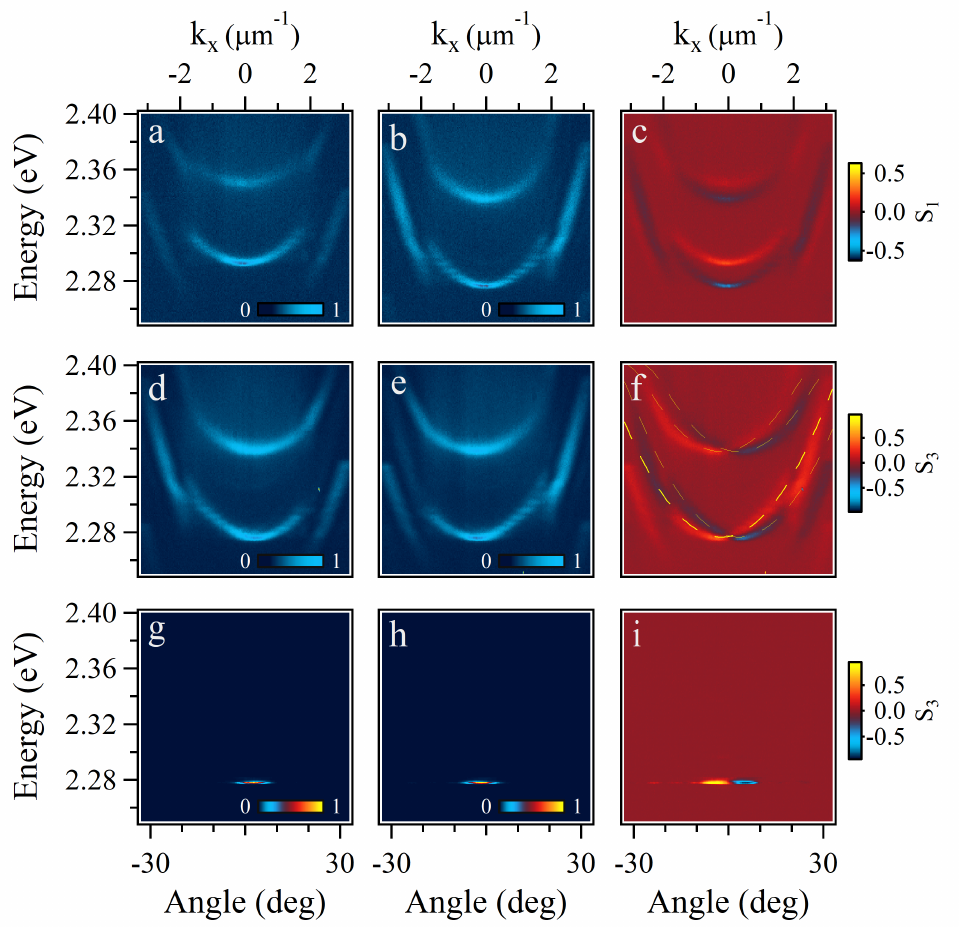} 
	\caption{\textbf{Rashba-Dresselhaus effect on polariton dispersion.} (a) Horizontally and (b) vertically linearly polarized dispersion below the threshold at 2 V. (c) Linearly polarized dispersion calculated from (a) and (b).
	(d) Left- and (e) right-circularly polarized dispersion below the threshold at 4.5 V. (g) Left- and (h) right-circularly polarized dispersion above the threshold at 4.5 V. (f, i) Circularly polarized dispersion corresponding to (d,e), and (g,h), respectively. The fitted curves in (f) are calculated by the Hamiltonian (1) with parameters: $E_\textup{x}=2.406$ eV, $E_\textup{p}$= 2.352 (2.284) eV for the upper (lower) set, $m=3.5\times10^{-6}m_\textup{e}$ ($m_\textup{e}$ is the free electron mass), $\xi=$ 5 meV, and $\beta_0=\beta_1=0$.} 
\end{figure}

Measurements discussed above were performed with a voltage of around 2 V applied across the microcavity, where the RD coefficients are too small such that the RD effect can be neglected. We increase the voltage in small steps and monitor the dispersion of the microcavity below and above the threshold. Increasing the voltage moves only the horizontally linearly polarized LP branches with the vertically linearly polarized branches fixed, such that the energy difference between the two sets is decreased, showing the same behavior as the cavity modes in \cite{19-liquid crystal_science}. When the two LP branches with opposite parity and orthogonal polarization come into resonance at the voltage of around 4.5 V, the Rashba interaction shifts the two branches apart along the $k_x$ direction. Fig. 3(d, e) shows the left- and right-handed circularly polarized dispersion along $k_x$ direction when $k_y$=0, where the two spin components are shifted towards opposite directions along $k_x$ (Fig. 3(f)). A similar spin-dependent shift of the dispersion is also observed above the threshold where two condensates with opposite parity and orthogonal polarization come into resonance, as shown in Fig. 3(g) and (h). We also measured the dispersion at fixed pumping fluence (below threshold, \textcolor{blue}{Supplementary Fig. 7(a-h)}, the pump fluence is the same as Fig. 3(d,e); above threshold,  \textcolor{blue}{Supplementary Fig. 7(i-p)}, the pump fluence is the same as Fig. 3(g,h)), but different voltages ranging from 3 V to 6 V, as shown in \textcolor{blue}{Supplementary Fig. 7}. One polariton branch is tuned to approach the other fixed branch when the voltage is increased from 3 V to 4.5 V. Further increasing the voltage splits the two polariton branches. During this process in \textcolor{blue}{Supplementary Fig. 7(i-p)}, the threshold for the condensation of the two polariton branches can vary greatly. For example, the population of the condensed polaritons in the lower-branch LP4 at the voltage larger than 5.5 V is reduced significantly, so that polaritons mainly condense to one branch LP5. In this case, the external voltage directly lets us manipulate the population and threshold of the polariton condensate, which is very challenging to be realized in common microcavities. 

\begin{figure}
	\centering
	\includegraphics[width=\linewidth]{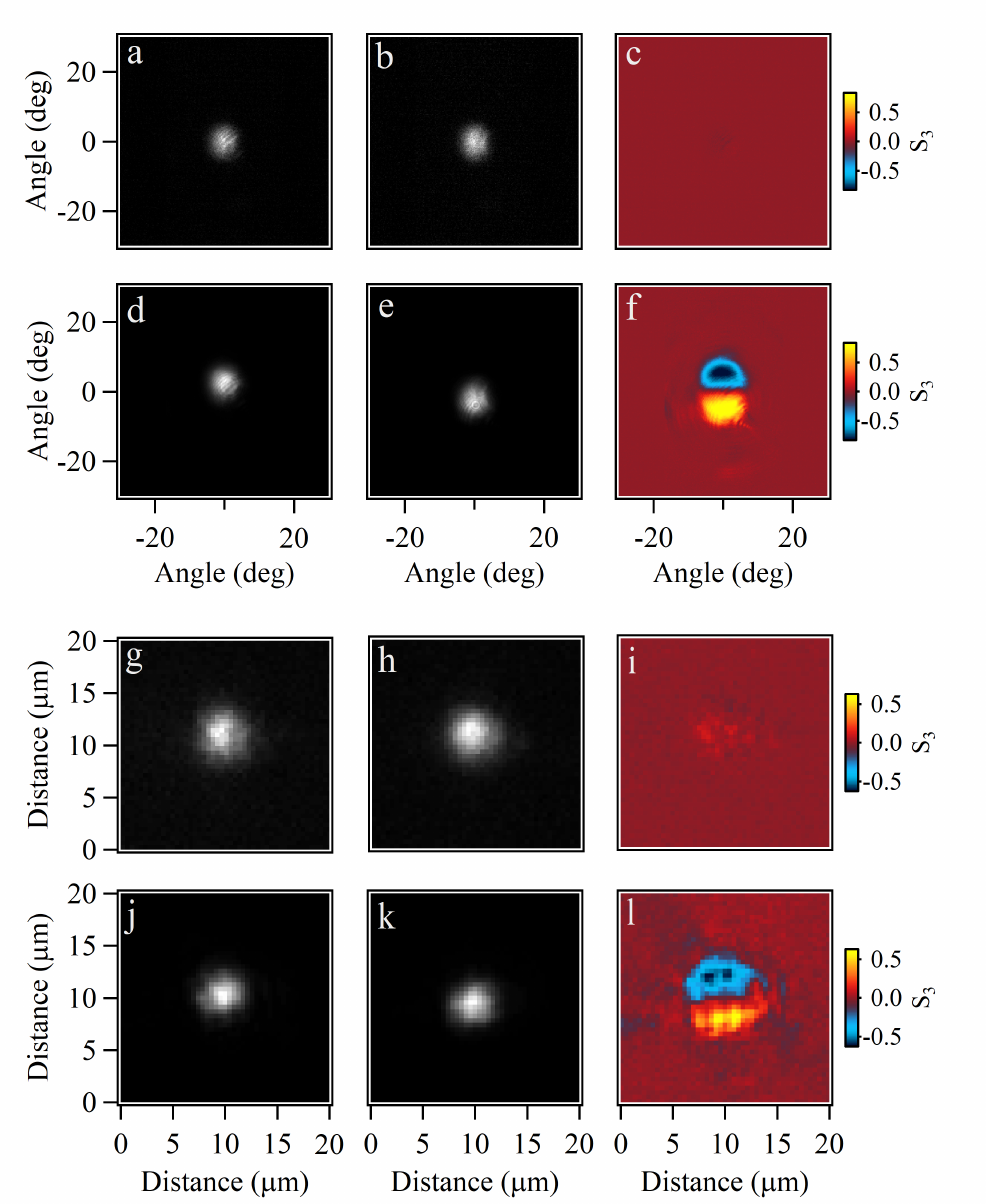} 
	\caption{\textbf{Rashba-Dresselhaus effect on polariton condensates in far-field (upper) and near-field (lower).} (a) Left- and (b) right-hand circularly polarized momentum space imaging at 0 V. (d) Left- and (e) right-hand circularly polarized momentum space imaging at 4.5 V. (c,f) Circularly polarized imaging corresponding to (a,b) and (d,e), respectively. (g-l) Corresponding real space imaging of the polariton condensate to (a-f), respectively.} 
\end{figure}

Finally we measure the spin dependent momentum and real space imaging of the polariton condensate under different voltage. In the experiments, the polaritons condense at $k_x=0$, such that we observe the condensate mainly located at the center of the Brillouin zone when the voltage is zero. The spin-dependent momentum space imaging can be seen in Fig. 4 where the spin-splitting of the polariton condensate is absent when the voltage is zero (Fig. 4(a-c)), whereas the polariton condensate is split at 4.5 V as shown in Fig. 4(d-f). In this scenario, the momentum of the polariton condensate is shifted and different spin components gain a non-zero momentum of around $k_x=\pm0.258\ \mu$m$^{-1}$. In accordance with the momentum space, the RD interaction in the microcavity modifies the distribution of the polariton condensate in real space due to the non-zero momentum gained. In the cold atom condensate, the two Raman lasers induce the coupling of the inner states of the atoms and create the Rashba interaction, which results in the separation of the atoms with different spins. Similar separation in real space with cavity photons is observed in \cite{poland spin separation}. In our experiments, we measure the distribution of the exciton polariton condensate in different spin polarizations and observe the spin-splitting of the polariton condensate in real space (see Fig. 4(j-l)) by about 0.86 $\mu$m. This spin-splitting of the polariton condensate in real space is also absent when the voltage across the microcavity is zero (Fig. 4(g-i)), demonstrating the RD effect occuring at 4.5 V. The other two Stokes parameters in momentum and real space are shown in \textcolor{blue}{Supplementary Fig. 8} and \textcolor{blue}{Supplementary Fig. 9}. These results clearly show the RD effect onto the exciton polariton condensates at room temperature.

Polaritons can also be trapped in a smaller perovskite microplate, leading to discrete energy levels (\textcolor{blue}{Supplementary Fig. 10}). To this end, we choose a smaller perovskite microplate but in the same microcavity to investigate the influence of the RD effect on higher-order polariton modes. From the results shown in Fig. 5, one can see that besides the ground state, the higher-order modes with opposite spins are also split by the RD effect and the circularly polarized momentum space imaging shows a particular spin texture. Thus, this kind of perovskite microcavity can also be used to explore novel non-Hermitian physics \cite{38Gao Nature,39Gao PRL}, involving remarkable and even complicated RD spin-splitting.

\begin{figure}
	\centering
	\includegraphics[width=\linewidth]{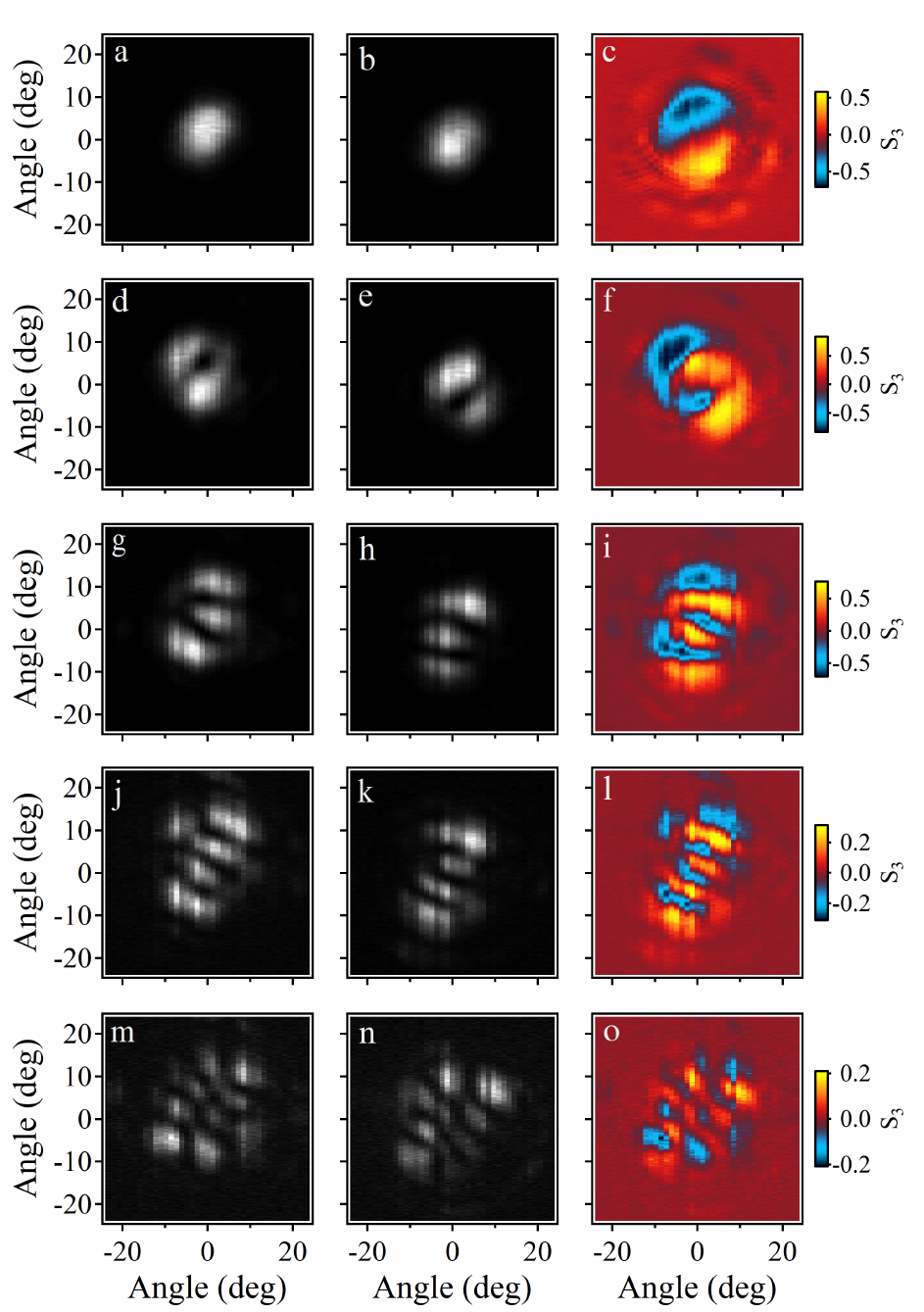}
	\caption{\textbf{Rashba-Dresselhaus effect on discrete higher-order modes of polariton condensates.} 
	(a, d, g, j, m) Left- and (b, e, h, k, n) right-hand circularly polarized momentum space imaging at different discrete polariton energy levels. (c, f, i, l, o) Circularly polarized momentum space imaging of each mode. The corresponding energies are 2.2772 eV, 2.2812 eV, 2.2846 eV, 2.2881 eV, and 2.2905 eV, respectively.}
\end{figure}

\section*{D\lowercase{iscussion}}

In this work we use a LC filled microcavity to demonstrate and  investigate the spin resolved  distribution  of a polariton condensate in real and momentum space at room temperature. The RD effect originating from the anisotropy of the LC microcavity splits the polariton condensate in both real and momentum space. Our work offers the possibility to study nontrivial band structures \cite{Lagoudakis} based on polariton condensates in a CsPbBr$_3$ microcavity at room temperature and the influence of the RD spin-splitting on nonlinear phenomena, such as solitary waves, optical bistability or multistability, modulational instability, and vortices \cite{Zhai hui, Ma xuekai}.

\section*{M\lowercase{ethods}}

\textbf{\textcolor{blue}{Fabrication} of CsPbBr$_3$.} Firstly, the Si/\textcolor{blue}{SiO$_2$} was ultrasonically cleaned with acetone, anhydrous ethanol and deionized water for 10 minutes, baked for 2 hours to completely dehydrate, and then put into the downstream of a quartz tube installed in a single temperature zone furnace. Secondly, the mixture of PbBr$_2$ and CsBr was put into the quartz boat, and then the quartz boat was put into the heating center of the single temperature zone furnace. The furnace was heated and was cooled naturally. Finally, the all-inorganic CsPbBr$_3$ is obtained. 

\textbf{\textcolor{blue}{Fabrication} of microcavity.} The DBR is fabricated through electron beam evaporation of alternating layers of silicon dioxide (SiO$_2$) and tantalum pentoxide(Ta$_2$O$_5$) (8 periods) with the thickness of 91 nm and 62 nm on glass substrate. The center wavelength and the band width of the DBR stop band are 530 nm and 160 nm, respectively.  The microcavity consists of two DBRs combined using UV curable adhesive (ergo8500).

\section*{D\lowercase{ata availability}}
All the data supporting this work 
are available from the corresponding author on reasonable request.

\section*{C\lowercase{ode availability}}
The code for the analysis of the data is available from the corresponding 
author on reasonable request.
\section*{R\lowercase{eferences}}

\section*{A\lowercase{cknowledgments}}
\begin{acknowledgments}
T.G. thanks for support from the National Natural Science Foundation of China (grant No. $11874278, 12174285$). The Paderborn group acknowledges funding from the Deutsche Forschungsgemeinschaft (DFG) through the collaborative research center TRR142 (project A04, No. 231447078), Heisenberg program (No. 270619725), and individual grant (No. 467358803). 
\end{acknowledgments}

\section*{A\lowercase{uthor contributions}}

T.G. and X.M. conceived the ideas and led the project, Y.L. performed the experiment and analyzed the results, X.M. performed the numerical simulation and theoretical analysis together with S.S.. T.G. and X.M. prepared the manuscript with contributions from M.G, H.D. and S.S.. Z.X fabricated the perovskite microplate. All authors discussed the results. \textbf{\textcolor{blue}{Corresponding authors: Xuekai Ma (xuekai.ma@gmail.com) and Tingge Gao (tinggegao@tju.edu.cn).}}

\section*{C\lowercase{ompeting interests}}
The authors declare no competing interests.

\section*{F\lowercase{igure Captions (for main text figures)}}

\noindent{\textbf{Figure 1: Schematics of the microcavity.} Schematic diagram of the liquid crystal microcavity structure with CsPbBr$_3$ inserted as the active layer. The anisotropy of the liquid crystal can induce spin orbit coupling of the polariton modes and split the polaritons with opposite spin, which is measured in the circularly polarized light emitted from the microcavity.}
\vspace*{\baselineskip} 
\newline
{\textbf{Figure 2: Polariton condensation in the LC microcavity.} (a) The photoluminescence dispersion of the microcavity below the threshold and (b) the condensate spectrum above the threshold for voltage of 2 V. (c) Integrated intensity and linewidth as  function of the pump fluence. (d) Polariton energy peak position as function of the pump fluence. \textcolor{blue}{The error bars in (c) are deviations of the estimated linewidth of polariton modes. And the error bars in (d) are deviations of the estimated energy of polariton modes.} The blue dashed lines in (a) are a guide for the eye to show the modes with large curvature. \textcolor{blue}{LP1-LP4 denote low branch polaritons 1-4.}}
\vspace*{\baselineskip} 
\newline
{\textbf{Figure 3: Rashba-Dresselhaus effect on polariton dispersion.} (a) Horizontally and (b) vertically linearly polarized dispersion below the threshold at 2 V. (c) Linearly polarized dispersion calculated from (a) and (b).
	(d) Left- and (e) right-circularly polarized dispersion below the threshold at 4.5 V. (g) Left- and (h) right-circularly polarized dispersion above the threshold at 4.5 V. (f, i) Circularly polarized dispersion corresponding to (d,e), and (g,h), respectively. The fitted curves in (f) are calculated by the Hamiltonian (1) with parameters: $E_\textup{x}=2.406$ eV, $E_\textup{p}$= 2.352 (2.284) eV for the upper (lower) set, $m=3.5\times10^{-6}m_\textup{e}$ ($m_\textup{e}$ is the free electron mass), $\xi=$ 5 meV, and $\beta_0=\beta_1=0$.} 
\vspace*{\baselineskip} 
\newline
{\textbf{Figure 4: Rashba-Dresselhaus effect on polariton condensates in far-field (upper) and near-field (lower).} (a) Left- and (b) right-hand circularly polarized momentum space imaging at 0 V. (d) Left- and (e) right-hand circularly polarized momentum space imaging at 4.5 V. (c,f) Circularly polarized imaging corresponding to (a,b) and (d,e), respectively. (g-l) Corresponding real space imaging of the polariton condensate to (a-f), respectively.} 
\vspace*{\baselineskip} 
\newline
{\textbf{Figure 5: Rashba-Dresselhaus effect on discrete higher-order modes of polariton condensates.} 
	(a, d, g, j, m) Left- and (b, e, h, k, n) right-hand circularly polarized momentum space imaging at different discrete polariton energy levels. (c, f, i, l, o) Circularly polarized momentum space imaging of each mode. The corresponding energies are 2.2772 eV, 2.2812 eV, 2.2846 eV, 2.2881 eV, and 2.2905 eV, respectively.}
\end{document}